\input harvmac
\overfullrule=0pt
\parindent 25pt
\tolerance=10000

\input epsf
\def\half{{\textstyle{1\over2}}}



\ifx\epsfbox\UnDeFiNeD\message{(NO epsf.tex, FIGURES WILL BE IGNORED)}
\def\figin#1{\vskip2in}
\else\message{(FIGURES WILL BE INCLUDED)}\def\figin#1{#1}\fi


\lref\witbub{E.~Witten,
 ``Instability Of The Kaluza-Klein Vacuum'',
Nucl.\ Phys.\  {\bf B195} (1982) 481.
}

\lref\motl{L. Motl, private communication.}

\lref\hort{F.~Dowker, J.~P.~Gauntlett, G.~W.~Gibbons and G.~T.~Horowitz,
``Nucleation of $P$-Branes and Fundamental Strings'',
Phys.\ Rev.\  {\bf D53} (1996) 7115
[hep-th/9512154].}

\lref\garstr{D.~Garfinkle and A.~Strominger,
``Semiclassical Wheeler wormhole production'',
Phys.\ Lett.\  {\bf B256} (1991) 146.
}

\lref\costa{M.~Costa and M.~Gutperle, ``The Kaluza-Klein Melvin Solution in
M-theory'', [hep-th/0012072].}

\lref\berg{O.~Bergman and M.~R.~Gaberdiel,
``Dualities of type 0 strings'',
JHEP {\bf 9907} (1999) 022
[hep-th/9906055].}

\lref\tse{J.~G.~Russo and A.~A.~Tseytlin,
``Magnetic flux tube models in superstring theory'',
Nucl.\ Phys.\  {\bf B461} (1996) 131
[hep-th/9508068].}

\lref\callan{C.~G.~Callan and J.~M.~Maldacena, `` Brane Dynamics from the
Born-Infeld Action'', Nucl. Phys. {\bf B513} (1998) 198, [hep-th/9708147].}

\lref\horava{M.~Fabinger and P.~Horava, ''Casimir Effect between
World-branes in herterotic M-theory'', [hep-th/0002073].}

\lref\savvidy{K.~G.~Savvidy,
``Brane death via Born-Infeld string,''
hep-th/9810163.
}

\lref\myersperry{R.~C.~Myers and M.~J.~Perry,
``Black Holes In Higher Dimensional Space-Times,''
Annals Phys.\ {\bf 172} (1986) 304.
}

\lref\galtsov{C.~Chen, D.~V.~Gal'tsov and S.~A.~Sharakin,
``Intersecting M-fluxbranes,''
Grav.\ Cosmol.\ {\bf 5} (1999) 45
[hep-th/9908132].
}

\lref\juan{J.~Maldacena,
``The large N limit of superconformal field theories and supergravity,''
Adv.\ Theor.\ Math.\ Phys.\  {\bf 2}, 231 (1998)
[hep-th/9711200].
}

\lref\melvin{M.~A.~Melvin, ``Pure Magnetic and Electric Geons'',
Phys. Lett. {\bf 8} (1964) 65.} 

\lref\avm{M.~Atiyah, J.~Maldacena and C.~Vafa,
``An M-theory flop as a large n duality,''
hep-th/0011256.
}
\lref\dowkera{F.~Dowker, J.~P.~Gauntlett, S.~B.~Giddings and G.~T.~Horowitz,
``On pair creation of extremal black holes and Kaluza-Klein monopoles'',
Phys.\ Rev.\  {\bf D50} (1994) 2662
[hep-th/9312172].}

\lref\dowkerb{F.~Dowker, J.~P.~Gauntlett, G.~W.~Gibbons and G.~T.~Horowitz,
``The Decay of magnetic fields in Kaluza-Klein theory,''
Phys.\ Rev.\ D {\bf 52} (1995) 6929
[hep-th/9507143].
}

\lref\gibbonsa{G.~W.~Gibbons,
``Quantized Flux Tubes In Einstein-Maxwell Theory And Noncompact
Internal Spaces'',
Print-86-0411 (CAMBRIDGE)
{\it Presented at 22nd Karpacz Winter School of Theoretical Physics: Fields
  and Geometry, Karpacz, Poland, Feb 17 - Mar 1, 1986}.}

\lref\garfb{D.~Garfinkle, S.~B.~Giddings and A.~Strominger,
``Entropy in black hole pair production'',
Phys.\ Rev.\  {\bf D49} (1994) 958
[gr-qc/9306023].
}

\lref\gibbonsb{G.~W.~Gibbons and K.~Maeda,
``Black Holes And Membranes In Higher Dimensional Theories With
Dilaton Fields'',
Nucl.\ Phys.\  {\bf B298} (1988) 741.
}

\lref\oz{O.~Aharony, S.~S.~Gubser, J.~Maldacena, H.~Ooguri and Y.~Oz,
``Large N field theories, string theory and gravity,''
Phys.\ Rept.\ {\bf 323} (2000) 183
[hep-th/9905111].
}

\lref\juanb{N.~Itzhaki, J.~M.~Maldacena, J.~Sonnenschein and S.~Yankielowicz,
``Supergravity and the large N limit of theories with sixteen  supercharges,''
Phys.\ Rev.\ D {\bf 58} (1998) 046004
[hep-th/9802042].
}

\lref\tseb{J.~G.~Russo and A.~A.~Tseytlin,
``Green-Schwarz superstring action in a curved magnetic Ramond-Ramond
background,'' 
JHEP{\bf 9804} (1998) 014
[hep-th/9804076].
}
\lref\nonsusa{I.~R.~Klebanov and A.~A.~Tseytlin,
``D-branes and dual gauge theories in type 0 strings,''
Nucl.\ Phys.\ B {\bf 546} (1999) 155
[hep-th/9811035].
}

\lref\nonsusb{A.~A.~Tseytlin and K.~Zarembo,
``Effective potential in non-supersymmetric SU(N) x SU(N) gauge theory
and interactions of type 0 D3-branes,'' 
Phys.\ Lett.\ B {\bf 457} (1999) 77
[hep-th/9902095].
}
 
\lref\nonsusc{I.~R.~Klebanov,
``Tachyon stabilization in the AdS/CFT correspondence,''
Phys.\ Lett.\ B {\bf 466} (1999) 166
[hep-th/9906220].
}

\lref\nonsusd{A.~M.~Polyakov,
``The wall of the cave,''
Int.\ J.\ Mod.\ Phys.\ A {\bf 14} (1999) 645
[hep-th/9809057].
}

\lref\nonsuse{A.~Adams and E.~Silverstein,''Closed String Tachyons,
AdS/CFT and Large N QCD'', [hep-th/0103220].}

\lref\saffin{P.~M.~Saffin,
``Gravitating fluxbranes,''
gr-qc/0104014.
}

\def\p{\partial}

\baselineskip 20pt plus 2pt minus 2pt

\Title{\vbox{\baselineskip12pt \hbox{hep-th/0104136}
\hbox{HUTP-01/A018}  }}
{\vbox{\centerline{Fluxbranes in String Theory}}}  
\centerline{Michael Gutperle and
Andrew Strominger   }
\bigskip\centerline{Jefferson Physical Laboratory}
\centerline{Harvard University}
\centerline{Cambridge, MA 02138}

\vskip .3in \centerline{\bf Abstract}

A flux p-brane in D dimensions has (p+1)-dimensional 
Poincare invariance and a nonzero rank 
(D-p-1) field strength 
tangent to  the transverse dimensions. We find a family of such solutions
in string theory and M-theory and investigate their  
properties. 

\noblackbox

\Date{April 2001}

\listtoc
\writetoc 

\newsec{Introduction}
  A magnetic monopole 
 is characterized by the integral of the field strength $F$ over the two
 sphere surrounding the monopole. A magnetic fluxtube, 
in contrast, is characterized by the integral of 
$F$ over the transverse plane. The  
much-studied p-branes of string theory are the generalizations 
of magnetic monopoles 
to higher rank field strengths and dimensions. The present work concerns 
the much less-studied generalizations of magnetic fluxtubes to higher
 rank and dimensions,  
which are referred to as fluxbranes or Fp-branes.\foot{ To be precise
an Fp-brane 
in D-dimensions has $ISO(p,1)\times SO(D-p-1)$ symmetry and a 
nonzero rank $(D-p-1)$ field strength, dual field strength, or wedge
 product of field strengths  
tangent to the transverse dimensions.} We shall find solutions of this 
type for all $p<8$ in string theory, as well as for $p=3,6$ in M-theory.  
These appear to be interesting family of 
(in general non-supersymmetric) solutions of string theory. 
Several duality relations involving the RR two-form field strength  
will be discussed. These are perhaps part of a larger web of 
dualities involving all rank forms and Fp-branes which remains to be found. 

The classic example of a fluxbrane with gravity is 
the Melvin universe \melvin, which is a F1-brane in 
3+1 Einstein-Maxwell gravity (and can 
be embedded in string theory). The solution is 
\eqn\ftylm{\eqalign{ds^2&=(1+{B^2r^2 \over 4})^2(-dt^2+dz^2+dr^2)
+{r^2\over (1+{B^2r^2 \over 4})^2 }d\phi^2, \cr
F&={Brdr\wedge d\phi\over (1+{B^2r^2 \over 4})^2 }.}}
$B$ here is the magnetic field strength along the axis 
$r=0$.  The total magnetic flux is 
\eqn\ftv{{1 \over 4 \pi}\int_{R^2}F={1 \over B}.}
This is finite, so magnetic flux is in a sense confined by gravity. 
At large $r$, the orbits of $\phi$ become small and the transverse space
resembles the surface of a teardrop with an infinite tail. 

In string theory one encounters dilatonic generalizations of this
solution \refs{\gibbonsa,\gibbonsb,\dowkera}.
The
simplest of these is the IIA F7-brane, which is considered in section 2. 
In 2.1 we review the M-theory description of the F7 which is 
simply an identification of flat
$R^{11}$ involving a rotation \refs{\dowkera,\dowkerb,\hort}. This
description strongly suggests a  
surprising periodicity in the field strength, as well as a dual relation
to 0A \refs{\costa,\berg} described in 2.2. 
This is used to motivate a conjecture in subsection 2.3 that the endpoint
of 0A tachyon condensation is the IIA vacuum. Intriguing similarities 
between this picture of 0A closed string tachyon condensation 
and $D-\bar D$ open string tachyon condensation are pointed out.
In section 2.4 we derive a 
dual relation of the F7-brane to the IIA theory on a cone by 
making a new choice for the M-theory circle. In section 3 we consider 
the supersymmetric IIA F5-brane which is characterized by a nonzero
integral of  
$F\wedge  F$ over the transverse $R^4$. This also has a flat M-theory 
lift. We show, by making a new choice of the
M-theory circle, that this is dual to IIA theory on an $A_N$ ALE space.

This beautiful duality story for the F7-brane 
was possible to uncover because of its simple M-theory lift. 
It is natural to expect that 
the other Fp-branes
(which are in some cases related by T-duality \galtsov) are also part of an 
as-yet-uncovered web of dualities. The rest of the paper
contains some preliminary investigations of the $p<7$ case.  
In section 4 we turn to the general Fp-brane solutions in M and string 
theory. These differ
qualitatively from the IIA F7 in that the flux is not confined and the 
integral of the field strength over the transverse dimensions does not
converge. The solutions can be found analytically only in the special 
case that the field strength $F$ at the origin goes to infinity 
(but is finite everywhere else). 
These analytic 
solutions have a warped cone-like structure with a singularity at the
origin. The singularity is resolved, but the asymptotic behavior
unchanged, by making $F$ finite at the origin. We find the solutions
 perturbatively both near the origin and 
at infinity and then match 
numerically. This is less difficult than it might sound because 
the asymptotic solution is a
universal attractor for all regular initial conditions at the origin. 
In section 5 we consider a special type of excitation of the
Fp-brane, namely static but unstable D(p-1)-brane bubbles. These are 
analogs of the open string dipoles in an electric field. We find a critical
scaling limit in which the string mass is taken to infinity while the
radius of and gauge coupling on the D-brane bubble is kept finite. 
Formally this is a field theory limit. In section 6 we speculate on a
possible holographic dual for the Fp-brane. In section 7 we ask whether or
not there could be flux periodicity for $p<7$.  Some evidence is provided
for the case of the Melvin universe
from the properties of black hole pair production.

\newsec{The IIA F7-brane}

In this section we discuss the IIA F7-brane. This case is characterized by 
a nonzero but finite  integral of the RR two 
form field strength over the two dimensions transverse 
to the brane. 
\subsec{The Solution }

The simplest way to describe the IIA F7-brane 
is in terms of its lift to $M$-theory \hort .
The reduction from 11 to 10 dimensions involves the
usual shift identification, accompanied by a spatial rotation\foot{A
similar construction involving a boost rather than a rotation leads to 
a time-dependent solution  with 
nonzero electric flux. The function $\Lambda$ determining the dilaton 
in 2.5 becomes of the form  
$1+E^2x^+x^-$ which has zeros along past and future
spacelike surfaces. } of
the angle $\phi$  in the plane transverse to the F7-brane
\eqn\iftb{\eqalign{x^{11}&\sim x^{11}+2\pi n_1R, \cr \phi & \sim
\phi+2\pi n_2+2\pi n_1 B R^2.}}
In other words the dimensional reduction
is performed along  orbits of the Killing vector $q=
\partial_{x_{11}}+BR\partial\phi$ \refs{\dowkerb,\hort}. The factors of $R$
in the second  
equation have been chosen for later convenience. In order to perform the
Kaluza-Klein reduction to ten dimensions it is convenient to
introduce the coordinate $\tilde \phi = \phi-BRx^{11}$ which is
canonically identified and constant along orbits of $q$,
\eqn\ift{\eqalign{x^{11}&\sim x^{11}+2\pi
n_1R, \cr \tilde\phi & \sim \tilde \phi+2\pi n_2.}} In terms of
$\tilde \phi$ the eleven dimensional metric is
\eqn\rftx{ds_{11}^2=\eta_{\mu\nu}dx^\mu dx^\nu +dr^2+r^2(d\tilde
\phi+BRdx^{11})^2 +(dx^{11})^2.} 
This can be expressed as a IIA solution using the formula
\eqn\tamt{ds^2_{11}=e^{4\phi /3}(dx^{11}+RA_\mu dx^\mu)^2+e^{-2\phi /3}
ds^2_{10}.}One finds
\eqn\fvv{\eqalign{ds^2_{10}&=\sqrt{\Lambda}(dr^2-dt^2+(dx^1)^2+\cdots+(dx^7)^2)
+{r^2d\tilde \phi^2 \over \sqrt{\Lambda}}, \cr 
    A_{\tilde \phi}&={B r^2 \over \Lambda}, \cr
     e^{4\phi \over 3} &=\Lambda \cr
     \Lambda & \equiv 1+B^2R^2r^2.}}
The parameter $B$ governing the size of the rotation in \iftb\ 
is hence identified as the strength of the magnetic field at the origin, 
and as usual $R={g_s \over M_s}$ is the ratio of the 
string coupling to
    the string mass.  
Note that since radius of
compactification (the coefficient of $(dx^{11})^2$) grows like
$r$, the IIA theory is strongly coupled at large $r$.

\subsec{0A$\leftrightarrow$IIA Duality}

In this subsection we recall an interesting duality 
conjectured in \costa. 

We first note that for $B={2 \over R^2}=2{M_s^2 \over g_s^2}$, one has a 
$4\pi $ rotation in \iftb, which is equivalent to no rotation at all. 
Therefore IIA with this critical magnetic field 
is dual to the IIA vacuum! Even more provocative is the 
case $B={M_s^2 \over g_s^2}$, for which the rotation in \iftb\ has no effect on
bosons but gives a minus sign for fermions. According to the conjecture 
of Bergmann and Gaberdiel \berg, M-theory compactification on $S^1$ with
twisted fermion boundary conditions is the 0A string theory at 
half the string coupling. 
This implies that IIA theory with $B={M_s^2 \over g_s^2}$ is dual 
to 0A string theory! More generally it was conjectured in \costa\ 
 that 
\eqn\liwth{ 
IIA \bigl(B,g_s \bigr) {\leftrightarrow} 0A\bigl( B-
{M_s^2 \over g_s^2}, {g_s \over 2}\bigr).
}
One of the main pieces of evidence for this duality 
comes from further $S^1$ compactification to 9 dimensions, followed 
by a ``9-11'' flip \refs{\berg,\costa}. This results in a perturbative 
duality relating 
IIA and 0A on twisted circles. Indeed the partition function is 
exactly computable \refs{\tse, \tseb, \costa} and smoothly interpolates
from IIA to 0A as a function of $B$. 

Of course ultimately the arguments in favor of this duality 
involve extrapolations to strong coupling which are unprotected 
by supersymmetry. As such they are much weaker than the 
arguments for supersymmetric dualities. Phase transitions 
could invalidate the picture. Nevertheless 
we feel the picture developed in \refs{\berg, \costa} hangs together 
quite well and is worth pursuing.  

   The periodic behavior of the magnetic flux appears almost trivial from 
the M-theory perspective since the flux is equivalent to a
rotation. However from the IIA perspective it is quite surprising. 
In this perspective a field strength is turned on, 
breaking both supersymmetry and Lorentz
invariance. Then at a large critical value of this field strength,  
Lorentz invariance is restored in new dual variables. At a yet higher field
strength, Lorentz invariance is again restored, together with
supersymmetry. The variables of this new IIA theory are related by a
complicated nonlocal transformation to those of the old IIA theory. 
The spectral flow relating the states of the two IIA theories mixes
perturbative and 
nonperturbative excitations.\foot{After compactification to 9 dimensions 
and a 9-11 flip, the spectral flow can be followed perturbatively \tse\costa.}

\subsec{The Fate of the Tachyon}
  
Recently there has been much discussion of open string tachyon
condensation. 
In this subsection we make a speculation about the endpoint of  
0A (closed string) 
tachyon condensation.

Let us first review the F7-brane instability at strong coupling, 
where the M-theory picture may be used \costa.  The 11-dimensional
euclidean Kerr
solution \myersperry\  then provides an instanton with the correct
boundary conditions.  
It has a negative mode and so represents an instability. However the
interpretation of this instability  depends on whether we use the 
0A or IIA interpretation of the theory. In the IIA interpretation, 
it represents  a brany analog of Schwinger pair production in which 
the magnetic field is damped by the creation Kaluza-Klein monopoles 
in a spherical shell.  In the 0A interpretation one has instead a 
so called ``bubble of nothing'' \witbub. It is quite challenging
 to understand the 
endpoint of the instability in the 0A picture.  From the IIA picture 
we expect the 
pair creation should continue until the magnetic field is completely
dissipated and one reverts to the IIA vacuum. 

At weak coupling, the IIA instability can be understood as the 
creation of zero-energy 
spherical D6-branes. At small radius, a spherical D6 
of course  has
positive energy. However at larger radius the energy becomes zero and 
then negative due to the potential energy in the magnetic field.  

It is natural to suppose that at weak coupling in the 0A picture the 
instability under discussion can be identified with that of the 0A
tachyon. This fits with the conjecture of  \berg\ that 
at a critical value of the coupling, the tachyon mass becomes positive. At
this value the perturbative instability becomes nonperturbative and may be
identified with the Kerr instanton instability.

This picture of closed string tachyon condensation, and its relation to 
a non-perturbative instability at strong coupling,  has an intriguing
similarity 
to that of open strings in the $Dp-\bar Dp$ system.  The tachyon is an open
string 
stretched between a $Dp$-brane and a $\bar Dp$-brane.  Pulling the 
$Dp-\bar Dp$ pair sufficiently far 
apart raises the mass of the tachyon above zero 
and eliminates the
perturbative instability. However it is replaced by a nonperturbative 
instability. This is described by an instanton which is a tube connecting
the $Dp-\bar Dp$ pair, and mediates decay into a ``bubble of nothingness' 
\refs{\callan\horava -\savvidy}.  
Similarly, the 0A theory  has a perturbative tachyon instability at 
weak coupling which is replaced by decay into bubbles of nothing when the
tachyon gets a positive mass.

\subsec{F7 $\leftrightarrow $ IIA Cone Duality}

Given an 11-dimensional geometry, a circle 
must be chosen to obtain a 10-dimensional description. 
The 0A-IIA duality can be viewed as different choices of 
this circle. In both cases the circle lies in the 
torus parameterized by $x^{11}$ and the angle $\phi$ in the 
plane transverse to the F7-brane, but they differ by a modular
transformation. More general 
alternate IIA descriptions of the theory, in the spirit of 
\avm,  can be 
obtained by a $SL(2,Z)$ transformation of the torus in \ift\ 
\eqn\ifct{\eqalign{{} \hat
x^{11}&= {a } x^{11}+bR \tilde\phi, \cr \hat \phi & = { c \over R}
x^{11}+d\tilde \phi, }} identified as \eqn\ifxt{\eqalign{\hat
x^{11}&\sim \hat x^{11}+2\pi n_1R, \cr \hat \phi & \sim \hat
\phi+2\pi n_2,}} where the integers $a,~b,~c,~d$ obey $ad-bc=1$.
One then finds that the flat 11-dimensional metric \rftx\ 
becomes 
\eqn\rftxb{ds_{11}^2=\eta_{\mu\nu}dx^\mu dx^\nu
+dr^2+r^2\bigl((a-bBR^2)d\hat \phi+ (dBR -{c \over R}) d \hat
x^{11} \bigr)^2 +(d~d\hat x^{11}-bRd\hat \phi)^2.} Consider the
case that the total magnetic flux is an integer $N$. This implies
\eqn\tyh{ BR^2={1 \over N}.} We may then choose
\eqn\dxxz{a=1,~~~~b=N-1,~~~~c=1,~~~~d=N,} which reduces \rftxb\ to
\eqn\rxtx{ds_{11}^2=\eta_{\mu\nu}dx^\mu dx^\nu +dr^2+{r^2 \over N^2}
(d\hat \phi)^2 +\bigr(Nd\hat x^{11}-(N-1)Rd\hat \phi\bigl)^2.}
This corresponds to a compactification to a locally flat
ten-dimensional spacetime with a $Z_N$ identification about the
8-plane $r=0$. The string coupling is \eqn\ftg{g_s=(NRM_p)^{3/2}.}
There is also a flat $U(1)$ connection $A_\phi ={N-1 \over N}$.

More generally one may consider the case of rational $B$ defined
by \eqn\ztyh{ BR^2={m \over N}} for integer $m.$ We may then take
\eqn\dxz{c=m,~~~~d=N~~~~Na-mb=1.}  There are infinitely many $a,b$
satisfying this relation and we will choose the smallest $b$.
which reduces \rftxb\ to \eqn\rdtx{ds_{11}^2=\eta_{\mu\nu}dx^\mu
dx^\nu +dr^2+{r^2 \over N^2} (d\hat \phi)^2 +\bigr(Nd\hat
x^{11}-bRd\hat \phi\bigl)^2.} This differs from \rxtx\ only in the
flat connection.

In conclusion a F7-brane is dual to IIA on a flat cone with 
RR flux at the origin. This duality will have a more familiar analog in
the context of the supersymmetric branes of the next section.

\newsec{ALE $\leftrightarrow$ Supersymmetric  F5-brane Duality}

 It is interesting
to consider the case where the spacetime rotation accompanying the
shift of $x^{11}$ involves four (rather than two) space dimensions
and lies within a $SU(2)_L$ subgroup of $SO(4)$.  Parameterizing $R^4$ in
terms of two complex coordinates
\eqn\rfourc{z_1=x_6+ix_7 = r \cos \theta e^{i(\phi+\psi)},\quad z_2 =
x_8+ix_9=r \sin \theta 
e^{i(\phi-\psi)}. }
The identifications are given by
\eqn\iftc{\eqalign{x^{11}&\sim x^{11}+2\pi n_1R, \cr \phi & \sim
\phi+2\pi n_2+2\pi n_1 B R^2.}}
Again the Killing vector used in the reduction is $q= \partial_{x_{11}}+ BR
\partial \phi$. 
Such flux tubes
were described in detail in \hort. The identification then
preserves half of the supersymmetries, given by spinors which are
invariant under $SU(2)_L$.
\eqn\spinrot{\epsilon  \to e^{ BR^2(\Gamma_{67}+ \Gamma_{89})}\epsilon.}
The flat eleven dimensional metric $ds^2=-dt^2+\cdots \mid dz_1\mid^2+\mid
dz_2\mid^2 + 
dx_{11}^2$ can be expressed in the following way
\eqn\ffmet{\eqalign{ds^2&= -dt^2 +dx_1^2+ \cdots + dx_5^2 + \Lambda\big
( dx_{11}+ RA_{\tilde\phi}d\tilde\phi+ 
RA_\psi d\psi)^2 +dr^2 + r^2 d\theta^2\cr &+ {r^2\over 16 \Lambda}
\Big(d\tilde\phi^2+ 
(1+{B^2R^2r^2\sin^22\theta\over 4})d\psi^2 + 2(1+
{B^2R^2r^2(1-\cos2\theta)\over 
4})d\tilde \phi d\psi\Big),}}
where 
\eqn\lamdef{\Lambda=e^{4/3\phi}= 1+ {B^2 R^2r^2\over 4}}
and 
\eqn\gaugedef{A_{\tilde \phi} ={B r^2 \over 4\Lambda},\quad A_\psi=  {B
r^2
\cos 2\theta \over 4\Lambda}.}
This field configuration represents  a flux
fivebrane. The gauge fields in ten dimensions have a nonzero second Chern
class
\eqn\fwedgf{\int F\wedge F = \oint_{S_3} A\wedge F = {8\pi^2 \over B^2R^2}.}

Note that the gauge field strength is not self-dual, however since the
dilaton
and the metric are nontrivial, they also appear in the supersymmetry 
transformation rules (given in the Einstein frame)
\eqn\susyiia{\eqalign{\delta \psi_\mu &= D_\mu \epsilon +{1\over 64} 
e^{3/4\phi} \big( \Gamma_{\mu}^{\nu\rho}-14 \delta_\mu^\nu
\Gamma^\rho\big)
 \Gamma_{11}\epsilon F_{\nu\rho},\cr
\delta\lambda &={1\over \sqrt{2}}\partial_\mu \phi \Gamma^\mu
\Gamma^{11}\epsilon+{2\over 16 \sqrt{2}} e^{3/4\phi} \Gamma^{\mu\nu}
\epsilon F_{\mu\nu},}}
making a 1/2 BPS configuration possible.

Proceeding as for the F7-brane in section 2  one finds in the case
$BR^2 =m/N$,   that 
this supersymmetric fluxbrane is dual to the $A_N$ ALE geometry in
IIA. There is also nonzero RR two-form flux at the origin. This can be 
seen by 
integrating the potential around one of the generators of $\pi_1=Z_N$ 
which gives $1 \over N$.

This construction has obvious generalization to Fp-branes of any p by 
simply choosing different rotations.

\newsec{The General Fp-brane Solution}
In this section we want to analyze the equations for $p$ fluxbranes in
M-theory and  string theory. A Fp-brane  will have $p+1$
dimensional Poincare invariance in the 'worldvolume' and $SO(q)$
rotational invariance in the $q=D-p-1$ transverse directions. 

There will be a non vanishing flux of a field strength
$F_q$ tangent to the 
transverse directions. This is contrasted with the usual BPS branes
which carry a charge measured by integrating
the field strength over a sphere surrounding the brane.
For notational convenience in this section 
we will use the number of transverse dimensions
$q$ to label the fluxbranes, rather than $p= D-q-1$.
 
 The nontrivial fields are the graviton, a $q$
form field strength $F_q$ and in the case of string theory a dilaton. The
action in the Einstein frame is given by
\eqn\aceint{S= {1\over  l_p^{D-2}}\int d^{D}x \sqrt{g}\big( R-{1\over
2}\partial \phi\partial 
\phi - {1\over 2\; q!} e^{a\phi} F_q^2\big).}
For M-theory we have $D=11$ and set $a=0$ and $\phi=0$, the
field strength $F_4$ has   either $q=4$ for a magnetic fluxbrane or the
dual $q=7$ 
for an electric fluxbrane. 

For type II string theory we have $D=10$.  There are two cases,
firstly when the field strength $F_q$ comes from the Ramond-Ramond  sector  
$q$
runs from $q=2$ to $q=8$ and the dilaton coupling is $a=1/2(5-q)$.
Note that in type IIA(B) we have (RR) Fp-branes with odd(even) $p$ which is in
contrast with Dp-branes which have even(odd) $p$. 
Secondly If the field strength (or its dual) is coming from the NS-NS
antisymmetric tensor $B_{\mu\nu}$ we have a NS-F6 brane  with $q=3$
and $a=-1$ and a NS F2-brane with $q=7$ and $a=+1$. In the
Einstein frame the RR and NS Fp branes are simply related by $a\to -a$. In
the following we will focus on the RR Fp-branes for type II.

Our ansatz for the metric is
\eqn\fmetricfd{ds^2 = e^{2A(r)}\big( -dt^2+dx_1^2+\cdots+dx_{D-q-1}^2\big)
+dr^2 
  +e^{2C(r)} dS_{q-1}^2.}
The equation of motion for the field strength  $d*F=0$ can be easily
  solved\foot{In the case of the IIB F4-brane, one has to impose the
  self duality of the five form field strength, $F_5 = fM_p (e^{-5A+4C}
  \epsilon_{r\alpha_1\cdots \alpha_4}+ \epsilon_{tx_1\cdots x_4})$} 
\eqn\fstrgth{F_q = fM_p e^{-(D-q)A+(q-1)C-a\phi} \epsilon_{r\alpha_1 \cdots
    \alpha_{q-1}}.}
The dimensionless 
constant $f$ measures the field strength at the origin. All functions
in
the ansatz only depend on the radial coordinate $r$, hence the
equations of motion can be derived from a one dimensional
Lagrangian $L=T-V$ 
where
the kinetic term is given by
\eqn\tdefine{T= e^{(D-q)A+(q-1)C}\Big( -{1\over
2}\phi'\phi'+(D-q)(D-q-1)A'A'+(q-1)(q-2)C'C'
+2(q-1)(D-q)A'C'\Big).}
The potential term comes from the field strength and curvature terms
\eqn\potdefine{V= -e^{(D-q)A+(q-1)C}\Big( (q-1)(q-2)e^{-2C}+
{f^2M_p^2\over 2}  e^{-2(D-q)A-a\phi}\Big).}
Together with a 'zero energy constraint', coming from the $R_{rr}$
component
of Einstein's equation. 
\eqn\zeroen{E=T+V=0.} 
\subsec{The Attractor Solutions}
Before entering into a detailed discussion of the solutions we 
highlight some salient features. The 
M-theory solutions have the 
asymptotic form (in appropriate coordinates) 
at large radius
\eqn\mtnj{ds^2 \sim r^m(-dt^2+(dx_1)^2+\cdots+(dx_p)^2)+dr^2+nr^2d\Omega_{q-1}^2,}
where $p+q= 10$. The field strength 
behaves 
in these coordinates as 
\eqn\gsl{F_q \sim sr^{q-2}\epsilon_q.}
$s$, $m$ and $n$ are positive constants which depend only on 
$q$, and $\epsilon_q$ is the transverse volume. In particular, 
although the field strength and metric are not flat, 
the asymptotic form of 
the solution has no memory of the field strength at the origin.

Corrections to \mtnj\ and \gsl\ are a power series in 
${ 1 \over rfM_p}$, where $f$ is the dimensionless field strength at the 
origin. The larger $f$ is, the closer we can get to the origin before
corrections 
become important. Very near the origin space is flat. 
The solution may be viewed as a bowl with a flat bottom. 
When the field strength $f$ is small, one must go far from the origin
before spacetime curvature becomes important, and the flat bottom is
large.

In string theory cases there is also a dilaton, and a symmetry 
under shifting the dilaton and rescaling $F$. This leads to one extra 
parameter in both the asymptotic and exact solutions, which can be taken to
be the constant value of the dilaton at the origin. Otherwise the situation is 
similar. There are attractor solutions of the form \mtnj-\gsl\ 
to which all solutions tend.  In M-theory there is a one parameter family 
of solutions for each q which tend to a unique attractor, while in string
theory there is a two parameter family of solutions which tend to a one
parameter family of attractors labeled by the string coupling. 
 
\subsec{The M-Theory F6-brane}
There are two fluxbranes in M-theory: the 'magnetic' Fp-brane with
$q=4$ where 
$F_4$
is non vanishing, and the 'electric' F3-brane with $q=7$ case
where the dual field strength
$F_7$ is non vanishing.
In this subsection we consider the F6-brane. 
Setting $D=11$ and $q=4$ 
in the Lagrangian one arrives at the following
equations of motion
  \eqn\eqofma{\eqalign{A''+7A'A'+3A'C'-{f^2M_p^2\over 6} e^{-14A}&=0,\cr
C''+3C'C'+7A'C'+{f^2M_p^2\over 3} e^{-14A}-2e^{-2C}&=0.}}
The field strength is then given by
\eqn\fsta{F_{r\theta\phi\psi}= fM_pe^{-7A+3C}}
Where $\theta,\phi,\psi$ denote angular coordinates on the transverse three
sphere. The zero energy constraint \zeroen\ becomes
\eqn\energy{E=7A'A'+7A'C'+C'C'-{f^2M_p^2\over 12} e^{-14A}-e^{-2C}=0.}
The boundary conditions at $r=0$ are determined by the fact that at the
center of the fluxbrane the metric becomes  flat. This implies that 
\eqn\chofv{C(r) = \ln r + B(r),}
Where both $A$ and $B$ go to constants as $r\to 0$. The behavior of the
solution near $r=0$ can then be determined by a power series expansion,
which depends on $a_0=A\mid_{r=0}$.
\eqn\powers{\eqalign{A(r)&= a_0 +{1\over 48}e^{-14a_0}(rfM_p)^2 -{1\over
2}
      {5\over  1152} e^{-28a_0}(rfM_p)^4 +{\cal O}(r^6),\cr
B(r)&= - {5\over 144} e^{-14a_0}(rfM_p)^2+ {1\over 2} {437\over 51840}
e^{-28a_0}(rfM_p)^4+{\cal O}(r^6).}}
Note that only even powers of $r$ appear in the power series. Without loss
of generality we can fix $a_0=0$.

We have been unable to find an analytic solution of the
equations \eqofma\ with the regular boundary conditions given
above. It seems impossible to decouple the 
equations because of the potential term $e^{-2C}$ in
\eqofma\ which is present when $q\neq 2$, i.e. for all cases but the Melvin
fluxtube. However we have been able to find an
exact solution which is singular at $r=0$:  
\eqn\resnum{\eqalign{A(r) &= {1\over 7} \ln(rfM_p)-{1\over 14}\ln(18/7),\cr
  C(r) &= \ln(r)  -{1\over 2} \ln(27/14).}}
\ From \fsta\ it follows that 
 the  behavior of the field strength is given by
\eqn\asymfst{F_{r\alpha_1\alpha_2\alpha_3} = \left( {2^4 7^2\over 3^7
    }\right)^{1/2} {r^2}.}
Fluxbrane solutions which are regular at the origin exist and can be studied
numerically. The result is that the solutions
approach \resnum\ asymptotically as $r\to \infty$. In addition it can
be shown that there is a three parameter family of linearized
perturbations around the solution \resnum, which all decay at least
as fast as $1/r$ when $r\to \infty$, so that \resnum-\asymfst\ is an 
attractor solution.

The flux $\Phi = \oint_{S^3} A_3$ inside a sphere of
radius $r$ in the transverse space grows
like  $\Phi\sim r^3$ as
$r\to \infty$. This means that unlike for the case of the Melvin fluxtube,
 the flux of 
F6-brane spreads out over the transverse space.

To compare with the BPS branes it is convenient to make a change of the
radial coordinate to bring the metric into 'isotropic' form. Up to an
overall numerical factor one finds
\eqn\siomet{ds^2 \sim (\tilde rf M_p)^{7/3}\Big({-dt^2 + dx_1^2+\cdots
dx_{5}^2+ d\tilde r^2 \over \tilde r^2}+ {8\over 21}
d\Omega_{S_3}^2\big).}

\subsec{The M-Theory F3-brane}
The F3-brane can be obtained by setting $q=7$, $D=11$ and $a=0$ 
in \aceint. The equations of
motion are 
\eqn\eqofmb{\eqalign{A''+4A'A'+6A'C'-{f^2M_p^2\over 3} e^{-8A}&=0,\cr
C''+6C'C'+4A'C'+{f^2M_p^2\over 6} e^{-8A}-5e^{-2C}&=0.}}
The zero energy constraint becomes
\eqn\energyb{E= A'A' +{5\over 2}C'C'+ 4A'C' -{f^2M_p^2\over 24}e^{-8A}-{5\over
  2}e^{-2C}=0.}
The field strength is given by
\eqn\fidlstd{F_{r\alpha_1\cdots \alpha_6}= fM_pe^{-4A+6C}.}
Where $\alpha_1,\cdots ,\alpha_6$ denote the coordinates on the transverse
six sphere. Repeating the analysis for the F6-brane given in the
previous section one finds that the behavior of the fluxbrane solution
near $r\to 0$ is  
\eqn\pwerb{\eqalign{A(x)&= a_0 +{1\over 42}e^{-8a_0}(rfM_p)^2 
-{1\over 2} {59\over
    15876} e^{-16a_0}(rfM_p)^4 +{\cal O}(r^6),\cr
B(x)&= - {5\over 504} e^{-8a_0}(rfM_p)^2+ {1\over 2} {1159\over 635040}
e^{-16a_0}(rfM_p)^4 +{\cal O}(r^6).}}
Using numerical and perturbation methods as in section 4.1 one can
    find good evidence that the regular fluxbrane solution will behave
    asymptotically as $r\to \infty$ like
\eqn\asumb{\eqalign{A(r) &= {1\over 4} \ln (rfM_p) -{1\over 8}
\ln(9/2)+{\cal O}(1/r),\cr C(r)&= \ln r -{1\over
    2}\ln(27/20)+{\cal O}(1/r).}} 
Using \fstrgth\ it is easy to see that  the flux will grow like $\Phi
= r^6$ as $r\to \infty$. In an 
isotropic coordinate system the asymptotic form of metric will be 
\eqn\siometb{ds^2 \sim (\tilde rf M_p)^{8/3}\Big({-dt^2 + dx_1^2+\cdots
dx_{3}^2+ d\tilde r^2 \over \tilde r^2}+ {5\over 12}
d\Omega_{S_6}^2\big).}
\subsec{Type II fluxbranes}
In type II  case there is  a dilaton in the action \aceint. Setting $D=10$
and the dilaton coupling  $a=1/2(5-q)$
in the Lagrangian  the  equations of motion in the Einstein frame become
\eqn\eqofh{\eqalign{\phi''+\phi'((10-q)A'+(q-1)C') -{1\over 4}(5-q)f^2M_s^2
e^{-2(10-q)A-\half(5-q)\phi}&=0,\cr
A''+A'((10-q)A'+(q-1)C')-{q-1\over 16 }f^2M_s^2
e^{-2(10-q)A-\half(5-q)\phi}&=0,\cr 
C''+C'((10-q)A'+(q-1)C')-{(q-2)}e^{-2C}+{9-q\over 16}f^2M_s^2
e^{-2(10-q)A-\half(5-q)\phi}&=0. }}
The zero energy constraint becomes
\eqn\tpiien{\eqalign{E&=-{1\over
2}\phi'\phi'+(10-q)(9-q)A'A'+(q-1)(q-2)C'C'
+2(q-1)(10-q)A'C'\cr
&-(q-1)(q-2)e^{-2C}
-{1\over 2} f^2M_s^2 e^{-2(10-q)A-\half(5-q)\phi}=0. }}
There is a simple integral of motion which sets
\eqn\intofm{\phi= 4{5-q\over q-1}A.}
Note that when $q=5$, i.e. for the type IIB F4-brane, the dilaton
is a constant and decouples.  
Eliminating $\phi$ from the equations \eqofh\ reduces the system to
\eqn\eqofmj{\eqalign{A''+A'((10-q)A'+(q-1)C')-{q-1\over 16 } (fM_s)^2
e^{-2{15+q\over q-1}A}&=0,\cr
C''+C'((10-q)A'+(q-1)C')-(q-2)e^{-2C}+{9-q\over 16} (fM_s)^2
e^{-2{15+q\over q-1}A} &=0, }}
Hence the complexity of the equations for the type II fluxbranes is the
same as the one for M-theory fluxbranes. Note that the Melvin fluxtube is
the type IIA F7-brane which has $q=2$. In this case the $e^{-2C}$ term
disappear \eqofmj\   and the equations can be solved exactly, yielding
the solution \fvv. 

 Making the same assumptions of
regularity at $r=0$ as for the M-fluxbranes implies that the solution
behaves near the origin as
\eqn\asymzer{\eqalign{A(r)&= a_0 + {q-1\over 32  q}e^{-2{15+q\over q-1}a_0}
(r f M_s)^2 + {\cal O}(r^4),\cr
B(r)&= {5-10q+q^2 \over 48 q(q-1)}e^{-2{15+q\over q-1}a_0} (r fM_s)^2
+
{\cal O}(r^4).} } 
Numerical  analysis  of the differential equations \eqofmj\ shows
that in the limit $r\to 
\infty$ the solution  behaves like
\eqn\asyminf{\eqalign{A(r)&= {q-1\over 15+q}\ln (r   fM_s) 
    -{q-1\over 2(15+q)}\ln\Big({128(3q-5)\over (15+q)^2}\Big) +{\cal
O}(1/r), \cr  
C(r) &= \ln (r) -{1\over 2}\ln\Big( {192(3q-5)\over (q-2)(15+q)^2}\Big)
+{\cal O}(1/r).}} 
The asymptotic behavior of the dilaton is then
\eqn\dilasym{e^{\phi}\sim  (r f M_s)^{4{(5-q)\over 15+q}}.}
This implies that the coupling blows up for $q<5$, i.e. for $p>4$
Fp-branes.
As $r\to \infty$ the form of the type II $q$ fluxbranes is given by
\eqn\asumc{ds^2\sim (rfM_s)^{2{q-1\over 15+q}}\left( {128(3q-5)\over
(15+q)^2}\right)^{-{q-1\over15+q}} (-dt^2+dx_1^2+\cdots+ dx^2_{9-q}) +
dr^2 + {(q-2)(15+q)^2\over 192(3q-5)} r^2 dS^2_{q-1}.}
In the new  coordinates the asymptotic metric in the Einstein
frame takes
the form 
\eqn\asymmet{ds^2 \sim\left({\tilde r  fM_s}\right)^{{15+q\over 8}}\big
( {-dt^2+dx_1^2+\cdots+ 
dx^2_{9-q} + d\tilde r^2 \over \tilde r^2}+{4\over 3}{(q-2)\over (3q-5)}
dS_{q-1}^2\big),} 
where we have dropped an overall numerical factor. The asymptotic behavior of
the dilaton is given by  
\eqn\asymdil{e^\phi \sim \left({\tilde r  f M_s}\right)^{5-q\over 4}.}
 To compare this with the near horizon geometry of $p$ branes, we have to
transform the metric to the string frame, i.e. multiply \asymmet\
 by $e^{\phi/2}$. Here we find a surprise: The conformal factor of the
metric does not depend on $q$:
\eqn\asymmetc{ds^2 \sim \left({\tilde r  f M_s}\right)^{5/2} \big
( {-dt^2+dx_1^2+\cdots+ 
dx^2_{9-q} + d\tilde r^2 \over \tilde r^2}+  {4\over 3}{(q-2)\over (3q-5)}
dS_{q-1}^2\big)} 
and the dilaton behavior is given by \asymdil.

\newsec{Critical D(p-1)-branes in Fp-branes}
   In this section we consider D-branes in a fluxbrane. In particular 
a D(p-1)-brane couples to the flux in a Fp-brane. We will describe  
a new scaling limit of D(p-1) branes in Fp-branes.  
\subsec{The Critical M2-brane in a F3-brane}

We begin with the simplest case which is the M2-brane.
The bosonic part of the action is
\eqn\drf{S_2=-M_p^3 \int d^3\sigma \sqrt{-\det( g_{\mu
\nu}\p_iX^\mu\p_jX^\nu)}+
M_p^3 \int {1\over 3!}C_{\mu \nu \rho}dX^\mu dX^\nu dX^\rho .}
We consider the  background
\eqn\bgk{g_{\mu \nu}=\eta_{\mu\nu},~~~~~~C_{0 \nu \rho}
=fM_p\epsilon_{0 \nu \rho \lambda}X^\lambda,}
where $f$ is dimensionless and $\epsilon$ is the volume element of a
flat (3+1)-dimensional subspace.
\bgk\ is a flat metric with a constant field strength
$F_4=dC=fM_p\epsilon$.
Of course such a field strength acts
as a source for metric curvature, so \bgk\ is valid only in a small
neighborhood of the origin. We will correct for this later when we consider
fluctuations of the membrane. 

Now consider a spherical M2-brane of radius $R$ oriented tangent to  
the flux $F_4=dC$. 
\bgk. The energy of such a brane is up to an overall numerical factor 
\eqn\egr{E(R)=M_p^3R^2-fM_p^4R^3.}
The first, positive,  term is just the total tension of the brane.
The second arises from the energy in the $F$ field. It is negative because
the field
strength is smaller in the interior of the spherical M2-brane (for the
chosen brane orientation).
The energy is stationary at
\eqn\rzro{R_0={2\over 3}{1\over  fM_p}.}
This corresponds to a static but  unstable M2-brane configuration.

This configuration has a simple analog. Consider letting go of  an
electron-positron pair in an electric field. If they are initially
far apart, their mutual attraction is negligible and they will be
accelerated further apart by the electric force. If they are
initially nearby, they will attract and annihilate.  In between
these two extremes is a critical radius at which they can remain
in unstable equilibrium. This configuration is the analog of the
unstable M2-brane. Note that both the M2-brane and the electron-positron
pair have no net charge.

A similar solution exists in euclidean space 
\eqn\bgke{g_{\mu \nu}=\delta_{\mu\nu},~~~~~~C_{\mu \nu \rho}
=fM_p\epsilon_{\mu \nu \rho \lambda}X^\lambda,}
with a $S^3$
(rather than $S^2\times R$) M2-brane. The critical radius for this solution is
\eqn\fji{R_0= {3 \over4}{1\over  fM_p}.} 
This is an instanton which describes decay of a constant
$F$ field through spherical brane creation \hort.
The branes are created with zero energy according to
\egr. They subsequently expand out to infinity, leaving a
dampened $F$ field in their interior. The brane worldvolume is the
deSitter space $dS_3$. This is the brane analog of
Schwinger pair production \hort.

Now we wish to consider a scaling limit in which we take
the Planck mass to infinity while holding the critical radius  \rzro\
(or equivalently \fji ) fixed:
\eqn\lmt{  M_p \to \infty,~~~~f\to 0~~~~ R_0={2 \over3}{1\over  fM_p}~~~{\rm
fixed}.}
Note that this requires a weak dimensionless field strength.
In order to compute the membrane field theory in this limit, 
the approximate solution \bgke\ is inadequate. We must use 
instead \pwerb. 
The  M2-brane action \drf\ 
near unstable equilibrium  for topology $S^2\times R$ 
can be written as
\eqn\drfsalt{\eqalign{S_2=-M_p^3 \int d^3\sigma\Big(
&\sqrt{-\det\big( \partial_i \rho\partial_j \rho + e^{2C(\rho)}
\eta_{\alpha\beta}\partial_i X^\alpha\partial_j X^\beta +
e^{2A(\rho)}( \partial_i R\partial_j R + {R^2\over R_0^2} h_{ij})\big)} \cr
+ \sqrt{-h}fM_p {R^3\over R_0^2}\Big).}}
In this expression
$h_{ij}$ is the radius $R_0$ (as given in \fji) metric on $S^2\times R$,
$X^m$ for $m,n=4,\cdots,10$ are fields in the 7 directions orthogonal
to the worldvolume directions of the fluxbrane, $\rho^2\equiv
\delta_{mn}X^mX^n$, $R$ is a field  tangent to 
the fluxbrane but orthogonal to the spherical  
brane, $A$ and $B$ are given in \pwerb\ with $a_0=0$ 
and the three longitudinal fields in \drf\ have been
eliminated by gauge fixing. Next we define fluctuations about the
static solution by \eqn\xddcx{R=R_0+{U \over M_p^{3/2}},
~~~~~X^m={U^m \over M_p^{3/2}}.} Near the scaling limit \lmt\ the
action then becomes \eqn\actl{S_2= -\int
d^3\sigma\sqrt{-h}\bigl({1\over 3}{M_p^3}+\delta_{mn}\nabla^j U^m
\nabla_j U^n+{2\over 7\dot 9}{ \delta_{mn} 
U^m  U^n \over R_0^2}
+\nabla^j U \nabla_j U -{U^2\over R_0^2}  \bigr) +{\cal O}({1\over
M_p^2}).}
 Note that the mass $m_U^2=-{1\over R_0^2}$ of the
tachyon $U$ is finite for $M_p \to \infty$. 

There is also a critical theory for 
a $dS_3$ brane with a $SO(3,1)$ symmetry and,
in euclidean 
 space,  with $SO(4)$
symmetry.  In the latter case one finds the euclidean version of
\actl\ but with $R_0$ given by \fji\ and where the mass of the
tachyonic excitation is $m^2_U= -3/(2R_0^2)$.

Of course the theory \actl\ is free for a single M2-brane.
The theory for $N$ M2-branes is strongly interacting at low energies but
poorly understood, even for a flat worldvolume. We will be more explicit
about the interacting theory for some D-brane cases.

\subsec{The Critical M5-brane in a F6-brane}

Let us now consider the analogous story for the M5-brane. The
action for the scalar fields is \eqn\drff{S_5=-M_p^6 \int
d^6\sigma \sqrt{-\det( g_{\mu \nu}\p_iX^\mu\p_jX^\nu)}+ M_p^6\int
{1\over 6!} \tilde C_{\mu_1\cdots \mu_6}dX^{\mu_1}\cdots dX^{\mu_6},} where
(when the 
Chern-Simons eleven-form can be ignored) $\tilde C$ is defined by
$*F=d\tilde C$. We consider the background 
\eqn\fbgk{g_{\mu
\nu}=\eta_{\mu\nu},~~~~~~\tilde C_{0\mu_2\cdots\mu_6}
=fM_p\epsilon_{0\mu_2\cdots\mu_7}X^{\mu_7}.} where $f$ is
dimensionless and $\epsilon$ here is the volume element of a flat
(6+1)-dimensional subspace. This gives a constant field strength
$F=*d\tilde C=fM_p*\epsilon$ which differs from the M2-brane case in that
it is purely spatial and has no time components. It corresponds to
the field strength of a F6-brane.

Now consider a spherical M5 of radius $R$ in the background
\fbgk. The energy of such a brane is
\eqn\sgr{E(R)=M_p^6R^5-fM_p^7R^6.}
The energy is stationary at
\eqn\rro{R_0={5 \over6}{1\over  fM_p}.}
This corresponds to a static but  unstable M5 configuration.
Holding this radius fixed with
$M_p
\to \infty$ 
yields the critical M5-brane theory. Defining \eqn\xddcx{R=R_0+{U
\over M_p^{3}}, ~~~~~X^m={U^m \over M_p^{3}}.} the action scales
\eqn\actlb{S_5= -\int
d^6\sigma\sqrt{-h}\bigl({1\over 6}{M_p^6}+\delta_{mn}\nabla^j U^m
\nabla_j U^n+{5^2\over 
2^5 3^2}{ \delta_{mn} 
U^m  U^n \over R_0^2}
+\nabla^j U \nabla_j U -{5\over 2}{U^2\over R_0^2}  \bigr) +{\cal O}({1\over
M_p^2}).}
In this expression $m,n$ run over 4 transverse dimensions.

In the euclidean version, one has a 
critical M5-brane solution with topology $S^6$ and radius
\eqn\gthl{R_0={6 \over7}{1\over  f M_p}.}
In the scaling limit the action has the same structure as \actlb\
where the mass of the tachyonic mode is $m^2_R= -3/R_0^2$ and the mass of the
transverse fluctuations is given by $m^2_{U^n}= 9/(2R_0^2)$.

In addition there are fermions and a self dual antisymmetric tensor field
$H=*H$ on the
worldvolume. This couples to the scalar fields through the
term $\int C\wedge H$ with $C$ as in \bgke. At distances short
compared to $R_0$ one recovers the $SO(7,1)\times SO(5)$ superconformally
invariant $(0,2)$ theory.

\subsec{The Critical Dp-brane in a  F(p+1)-brane}

The case of D-branes is only slightly more complicated. 
Consider
the DBI-WZW action for a Dp-brane 
\eqn\drft{\eqalign{S_p&=-{M_s^{p+1} \over
g_s}\int d^{p+1}\sigma e^{-\phi}\sqrt{-\det( g_{\mu \nu}\p_iX^\mu\p_jX^\nu
+{2\pi \over M_s^2}F_{ij})}\cr
&+ M_s^{p+1} \int{1\over (p+1)!} C_{\mu_1\cdots\mu_{p+1} }dX^{\mu_1}
\cdots dX^{\mu_{p+1}} ,}}
in the F(p+1)-brane background of section 
4. Near the origin we have the  background
\eqn\bsgk{g_{\mu 
\nu}=\eta_{\mu\nu},~~~~~~C_{0\mu_{2}\cdots \mu_{p+1}}=fM_s 
\epsilon_{0{\mu_2}\cdots\mu_{p+2}}X^{\mu_{p+2}},} where $f$ is
dimensionless and 
$\epsilon$ is the volume element of a flat (p+2)-dimensional
subspace. There is an unstable $S^p\times R$
brane solution at the critical radius \eqn\fty{R_0={p \over p+1}{1\over 
fg_sM_s}={p \over p+1}{1\over 
fg^2_{YM}M_s^{p-2}} , } 
where 
$g_{YM}^2=g_sM_s^{3-p}$ is the D-brane gauge coupling.   
Now we wish to take a scaling limit 
\eqn\sclm{M_s \to \infty, ~ ~~~~~~~~~R_0, ~~g_{YM}^2~~{\rm fixed}.}
The dimensionless field strength $f$ behaves as 
\eqn\jik{f={p \over p+1}{M_s^{2-p}\over
R_0g^2_{YM}} ,} 
which goes to zero or infinity depending on the value of $p$. 
As before we introduce scaling variables for the transverse fluctuations
\eqn\ftsl{U^m=M_s^2X^m, ~~m=1,\cdots, 8-p,~~~~~~U=M_s^2R,}
while the D-brane gauge field is not scaled. 
The power of $M_s^2$ insures that the kinetic terms remain finite as 
$M_s \to \infty$, as well as the (tachyonic) mass term for $U$,
\eqn\ttwotac{m^2_U= -{p\over 2}{1\over g_{YM}^2 R_0^2}. } 
However the mass $m_T$ for the fields $U^m$ transverse to 
the Fp-brane arise from quadratic corrections to the metric
and dilaton given in \asymzer\ where the metric has to be transformed
to the string frame.  These are a power series in $fM_sX^m$ which scales as 
\eqn\rdcyx{fM_sX^m \sim {M_s^{1-p} U^m \over R_0 g_{YM}^2}.}
Since the entire action is multiplied by $M_s^4 \over g_{YM}^2$, 
the mass goes as  
\eqn\rdx{m_{U^n}^2 = {5 p^2\over 8(9-p)(p+1)^2} {M_s^{6-2p} \over
R_0^2g_{YM}^6}.} 
Hence for $p>3$, the mass scales to zero. For $p=3$ it is finite but higher
order
corrections scale to zero. 
If $p<3$, $m_T^2$  scales to infinity and the fields $U^m$ are frozen.

 The resulting critical theory is similar to \actl\ but of course also has
$U(N)$ gauge fields (for the case of $N$ coincident branes). 
In the case of $p=3$ 
at distances
short compared to $R_0$, one has the superconformal ${\cal N}=4$
Yang-Mills. At scales of order $R_0$ the superconformal symmetry
is softly broken by for example a tachyonic mass term for one of
the six scalars.

\newsec{A Holographic Dual?}
It is natural to conjecture, in the spirit of
AdS/CFT \juan,   that the F-branes in
string theory and $M$-theory are the large $N$ duals of
field theories without gravity. This would be of special interest because
the duals should be nonsupersymmetric field theories. Several attempts to 
understand nonsupersymmetric holographic dualities
\refs{\nonsusa\nonsusb\nonsusc\nonsusd-\nonsuse}
have proceeded by looking for spacetime
solutions corresponding to nonsupersymmetric brane configurations. 
These have in general turned out to be singular at the origin, which has
hampered progress. Here we are taking the opposite tack: we are beginning
form a smooth spacetime solution and trying to find a field theory dual. 
Of course any such duality 
conjecture will be difficult to verify because
there is no supersymmetry and the solutions are unstable. 
In this section we will mention several possibilities.

Recalling the type II solutions of subsection 4.3, at 
 large radius the Fp-brane dilaton and metric are
\eqn\asymdil{e^\phi = \left({\tilde r  f M_s}\right)^{p-4\over 4},}
\eqn\asymmetb{ds^2 = \left({\tilde r  f M_s}\right)^{5/2} \big
( {-dt^2+dx_1^2+\cdots+ 
dx^2_{p} + d\tilde r^2 \over \tilde r^2}+  {4\over 3}{(7-p)\over (22-3p)}
dS_{8-p}^2\big).} 
The solution is the same in form as the near horizon D-brane solutions, 
but the coefficients differ. The fact that the blueshift factor 
$g_{tt} \to \infty$ as 
$r \to \infty$ suggests that $r=\infty$ corresponds to the UV. 

What could the dual field theory be? One possibility that comes to mind is 
the field theory of N non-BPS D-branes. Like the Fp-branes, these exist 
only for odd (even) p in type IIA (IIB) string theory. However there are a
number of question to which we have no answer, including: 
What happens to the tachyon on the non-BPS brane? 
What is N dual to? 
What is the origin of the RR flux? 
Why does the dilaton blow up only for $p>4$, since field theories tend to
be well behaved in the UV only for $p<4$? 

Another possibility, which addresses some of these questions but raises
others, is the critical D-brane bubbles of the previous section. The
behavior of the dilaton suggests that p is the dimension of the dual field
theory plus one.  This matches with the fact that the critical brane has
dimension one less than the Fp-brane in which it lies. It also is naturally
associated with RR flux. However a problem with this idea is that the 
field theory on the critical D3-brane in the F4-brane is N=4 Yang-Mills in
the UV. Hence we would expect the holographic dual to be asymptotic to 
$AdS_5\times S^5$. This does not seem to be the case. 

A further problem with this proposal is
that these bubbles are not the conformal boundaries of the 
fluxbranes. In the $AdS^5\times S^5$ example, the branes live on the 
$S^3\times R$ conformal boundary of the spacetime. (Of course it is not
clear which features of $AdS$ holography should carry over.) The
conformal boundary of the metric \asymmetb\ at $r=\infty$ is 
an $8-p$ sphere. This suggests the possibility of a dual field theory on 
$S^{8-p}$, but we have no concrete proposal for what that might be.  

It is also of interest to consider a dual DLCQ matrix description 
of the F7-brane. Interesting progress in the direction has recently 
been made in \motl. 

In conclusion holographic duality for Fp-branes remains an interesting and
open problem.

\newsec{Flux Periodicity}

The IIA F7-brane has a self-duality under shifts of $2\pi$ in the 
magnitude of the two-from field strength at the origin. 
This follows from its presentation as a twisted M-theory 
compactification. Although there is no simple M-argument, 
it is natural to expect that the other fluxbranes also 
have a periodic character. Indirect evidence for this follows from 
T-duality. A longitudinal T-duality of the IIA F7-brane gives 
the IIB F6-brane, smeared along the direction of the T-duality. 
This implies that the smeared -- and possibly also the unsmeared -- 
F6 is periodic (and is dual to 0B at 
a critical field strength). 
Further T-duality similarly suggests that all the fluxbranes 
may be periodic.

Further indirect evidence for periodicity can be found from 
examination of the flux decay rate. The IIA F7-brane decays by 
production of spherical D6-branes. In the M-theory picture these 
are Kaluza-Klein monopoles. Naively one expects the decay rate to grow 
with the field strength. However the instanton action goes to infinity at 
the critical field $B={2 \over R^2}$ implying there is no decay. 
This is consistent with the dual reinterpretation of this critical field
as 
the flat IIA vacuum. 

A similar phenomenon occurs in 
original non-periodic four-dimensional Melvin universe. This 
is a solution of $N=2$ supergravity with no vector multiplets, which can 
embedded in string theory for example by IIB compactification on a rigid 
Calabi-Yau. This Melvin universe decays by pair production of charged
black
holes rather than Kaluza-Klein monopoles. The instanton action for 
production of a pair of black holes with integer magnetic charges $\pm q$
in a magnetic
field $B$ 
is \refs{\garstr,\garfb} 
\eqn\fda{S=4\pi q^2{(1-Bq^2)\over 1-(1-Bq)^4}.}
The action goes to infinity, and the decay rate to zero, 
when 
\eqn\fffvvc{B={2 \over q}.}
The reason for this is that the black holes are pair created at a 
separation of order $1 \over B$. When $B$ gets to large, the black hole 
horizons touch and the pair cannot be pulled apart. At the critical field 
$B=2$ all pair production ceases. This suggests that 
we have reached a new supersymmetric vacuum - possibly just a dual
representation of the $B=0$ vacuum. 

\medskip
{\bf Note added:}
\medskip
After this work was completed, \saffin\ appeared in which  the singular 
 fluxbrane solutions presented in section 4 were found independently.
\medskip
{\bf Acknowledgements}
\medskip
We are grateful to 
Oren Bergman, Miguel Costa, Rajesh Gopakumar, Gary Horowitz, 
Shiraz Minwalla, Lubos Motl, 
Eva Silverstein and Cumrun Vafa for useful discussions. 
The work of M.G. is supported in part by the David and Lucile Packard
Foundation and that of A.S. by DOE grant DE-FG02-91ER40654.

\listrefs

\end